\def\year{2021}\relax

\documentclass[10pt,twocolumn,letterpaper]{article}

\usepackage
[
        letterpaper,
        left=.75in,
        right=.75in,
        top=.75in,
        bottom=.75in,
]{geometry}
\usepackage{authblk}

\date{}

\pdfoutput=1
\usepackage{times}  
\usepackage{helvet} 
\usepackage{courier}  
\usepackage[hyphens]{url}  
\usepackage{graphicx} 
\usepackage{subfig}
\usepackage{amssymb}
\usepackage{amsmath}
\usepackage{tabularx}
\usepackage{pdfpages}
\usepackage{graphicx}
\usepackage{tabularx}
\usepackage{enumitem}
\usepackage{dblfloatfix}
\usepackage{booktabs}
\usepackage{multirow}
\usepackage{ulem}
\usepackage{tikz}
\usepackage{pgfplots}
\usepackage[numbers]{natbib}  
\usepackage{caption} 
\newcommand{\ignore}[1]{}
\newcommand{\nop}[1]{}

\newcommand*{\ie}{\textit{i.e.}}
\newcommand*{\cf}{\textit{c.f.}}

\selectcolormodel{cmyk}

\def\addlegendimage{\csname pgfplots@addlegendimage\endcsname}

\pgfplotsset{every axis/.append style={
                    xlabel={$x$},          
                    ylabel={$y$},          
                    label style={font=\sffamily\scriptsize},
                    tick label style={font=\sffamily\scriptsize}  
                    },
                    legend image code/.code={
                \draw[mark repeat=2,mark phase=2]
                plot coordinates {
                (0cm,0cm)
                (0.15cm,0cm)        
                (0.3cm,0cm)         
                };%
                }
                    }
\pgfplotsset{compat=newest}

\begin{document}

\tikzstyle{textnode}=[fill=blue!15, align=left, draw=black, outer sep=2pt, inner sep=2pt, rounded corners, text width=3.0cm, minimum height=1.0cm, minimum width=3cm]
\tikzstyle{responsenode}=[fill=gray!5, align=left, draw=black, outer sep=2pt, inner sep=2pt, rounded corners, minimum height=1.0cm, text width=3.0cm, minimum width=2cm]
\tikzstyle{title}=[font=\large]
\tikzstyle{postedge}=[- >, thick, draw=blue]
\tikzstyle{commentedge}=[- >, thick, draw=red]
\tikzstyle{genericedge}=[- >, thick]
\tikzstyle{genericnode}=[fill=green!10, draw=black, rounded corners, outer sep=2pt, inner sep=2pt]
\tikzstyle{examplepostnode}=[fill=blue!15, align=left, draw=black, outer sep=2pt, inner sep=2pt, rounded corners, text width=4.5cm, minimum size=1.0cm]
\tikzstyle{exampleresponsenode}=[fill=gray!5, align=left, draw=black, outer sep=2pt, inner sep=2pt, rounded corners, minimum height=1.0cm, text width=4.5cm, minimum width=3cm]

\title{Analysis of Moral Judgement on Reddit}

\author{Nicholas Botzer, Shawn Gu, and Tim Weninger \\ Department of Computer Science and Engineering \\ University of Notre Dame
\\ \{nbotzer, sgu3, tweninger\}@nd.edu}


\maketitle

\begin{abstract}
    Moral outrage has become synonymous with social media in recent years. However, the preponderance of academic analysis on social media websites has focused on hate speech and misinformation. This paper focuses on analyzing moral judgements rendered on social media by capturing the moral judgements that are passed in the subreddit /r/AmITheAsshole on Reddit. Using the labels associated with each judgement we train a classifier that can take a comment and determine whether it judges the user who made the original post to have positive or negative moral valence. Then, we use this classifier to investigate an assortment of website traits surrounding moral judgements in ten other subreddits, including where negative moral users like to post and their posting patterns. Our findings also indicate that posts that are judged in a positive manner will score higher. 
\end{abstract}
\section{Introduction}
How do people render moral judgements of others? 
This question has been pondered for millennia. Aristotle, for example, considered morality in relation to the end or purpose for which a thing exists. Kant insisted that one's duty was paramount in determining what course of action might be good. Consequentialists argue that actions must be evaluated in relation to their effectiveness in bringing about a perceived good. 
Regardless of the particular ethical frame that one ascribes to, the common practice of evaluating others' behavior in moral terms is widely regarded as important for the well-being of a community. Indeed, ethnographers and sociologists have documented how these kinds of moral judgements actually increase cooperation within a community by punishing those who commit wrongdoings and informing them of what they did wrong \cite{Boyd_Richerson_1992}. 

The process of rendering moral judgement has taken an interesting turn in the current era with the adoption of the Internet and social media in particular. Online social systems allow people to encounter and consider the lives of others from around the world.
At no other time in history have so many people been able to examine  such a variety of cultures and viewpoints so readily. This increased sharing and mixing of viewpoints inevitably leads to online debates about various topics \cite{yardi_dynamic_2010}.
The content of these debates provides researchers with the opportunity to ask specific questions about argument, disagreement, moral evaluation, and judgement with the aid of new computational tools. 

To that end, recent work has resulted in the creation of statistical models that can understand moral sentiment in text \cite{Sagi_Dehghani_2014}.
However, these models rely heavily on a gazette of words and topics and their alignment on moral axes.
The central motivation for these works are grounded in moral foundation theory \cite{Graham_Haidt_Koleva_Motyl_Iyer_Wojcik_Ditto_2013} where studies also tend to investigate the use of morality as related to current events in the news such as politics or religion.
Despite their usefulness in understanding the moral valence of specific current events, the goal of the current work is to study moral judgements rendered on social media that apply to more common personal situations.

\begin{figure}[t]
 \centering    

    \subfloat[]{\includegraphics[width=0.43\textwidth]{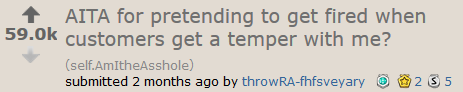}}

     \subfloat[]{\includegraphics[width=0.43\textwidth]{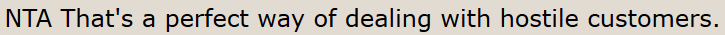}}

    \vspace{-0.25cm}
 \caption{\label{fig:aita-ex} Example of (a) a post title and (b) a comment in the {/r/AmItheAsshole} subreddit. The {NTA} prefix and comment-score (not shown) indicates that the commenter judged the poster as ``Not the Asshole''.}
\end{figure}

We focus on Reddit in particular, where users can create posts and have discussions in threaded comment-sections. 
Although the details are complicated, users also perform curation of posts and comments through upvotes and downvotes based on their preference~\cite{gilbert2013widespread,glenski2017rating}.
This assigns each post and comment a score reflecting how others feel about the content.
Within Reddit there are a large number of subreddits, which are small communities dedicated to various topics.
The subreddit of interest for our question regarding moral judgements is called {/r/AmItheAsshole}.
Users of this subreddit post a description of a situation that they were involved in; they are also encouraged to explain details of the people involved and the final outcome of the situation.
Posters to {/r/AmItheAsshole} are typically looking to hear from other Reddit users whether or not they handled their personal situation in an ethically appropriate manner.
Other users then respond to the initial post with a moral judgement as to whether the original user was an asshole or not the asshole. 
Figure \ref{fig:aita-ex} shows an example of a typical post and one of its top responses.
One important rule of {/r/AmItheAsshole} is that top-level responses must categorize the behavior described in the original post to one of four categories: Not the Asshole (NTA), You're the Asshole (YTA), No assholes here (NAH), Everyone sucks here (ESH).
In addition to providing a categorical moral judgement, the responding user must also provide an explanation as to why they selected that choice.
Reddit's integrated voting system then allows other users to individually rate the judgements with which they most agree (upvote) or disagree (downvote). After some time has passed the competition among different judgements will settle, and one of the judgements will be rated highest. This top comment is then accepted as the judgement of the community.
This process of passing and rating moral judgement provides a unique view into our original question about how people make moral judgements.

Compared to other methodologies of computational evaluation of moral sentiments, collecting judgements from {/r/AmItheAsshole} (AITA) has some important benefits.
First, because posters and commenters are anonymous on Reddit, they are more likely to share their sensitive stories and frank judgements without fear of reprisal~\cite{ong2000impact, jordan2020signaling}.
Second, the voting mechanism of Reddit allows a large number of users to engage in an aggregated judgement in response to the original post ~\cite{Glenski_Pennycuff_Weninger_2017}. However, the breadth and variety of this data does pose additional challenges. For instance, judgements are provided without an explicit moral-framing, and, similarly, Reddit-votes do not explicitly denote moral valence and are susceptible to path dependency effects~\cite{glenski2018guessthekarma}.

In the present work we use data from AITA to investigate how users provide moral judgements of others. We then extract representative judgement-labels from each comment and use these labels and comments to train a classifier. This classifier is then broadly applied to infer the moral valence of other Reddit comments from ten different subreddits and used to answer the following research questions:

\begin{itemize}[leftmargin=.5in]
    \item [RQ1:] What language is most closely associated with positive and negative moral valence?
    \item [RQ2:] Is moral valence correlated with the score of a post?
    \item [RQ3:] Do certain subreddit-communities attract users whose posts are typically classified by more negative or positive moral judgements?
    \item [RQ4:] Are self-reported gender and age descriptions associated with positive or negative moral judgements?
\end{itemize}

In summary, we find that posts that are judged to have positive moral valence (\ie, NTA label) typically score higher than posts with negative moral valence. We also find that certain subreddit-communities where users confess to something immoral (\ie, such as /r/confessions) tend to attract users whose posts are characterized by negative moral valence. Among these immoral users we show that their posting habits tend towards three different types. Finally, we show that self-described male-users are more likely to be judged an asshole than female-users.



\section{Methodology}\label{sec:data}
We retrieve moral judgements by collecting posts and comments from the subreddit /r/AmItheAsshole, taken from the Pushshift data repository \cite{Baumgartner_Zannettou_Keegan_Squire_Blackburn_2020}.

The questions raised in the present work are considered human subjects research, and relevant ethical consideration are present. We sought and received research approval from the Institution Review Board at the University of Notre Dame under protocol \#20-01-5751.

\begin{table}[t]
\centering
\begin{tabular}{@{}llr@{}}
    \toprule 
\textbf{Label} & \textbf{Meaning}      & \textbf{\# Comments}       \\ \midrule
NTA          & Not the Asshole     & 717,006\\ 
YTA          & You're the Asshole  & 372,850\\ 
NAH          & No assholes here   & 91,903 \\ 
ESH          & Everyone sucks here & 79,059\\ \bottomrule
\end{tabular}
\caption{The four judgements that users can pass on the subreddit /r/AmItheAsshole.}
\label{tab:abbreviations}
\end{table}

We restricted our data collections to posts submitted between January 1, 2017, and August 31, 2019. In order to assure that labels reflected the result of robust discussion we excluded those posts containing fewer than 50 comments. Subreddit rules require that top-level comments begin with one of four possible prefix-labels indicated in Tab.~\ref{tab:abbreviations}. Because of this rule, we further restrict our data collection to contain only top-level comments and their prefix-label. Comments with the INFO prefix, which indicates a request for more information, and comments with no prefix are also removed from consideration. This methodology resulted in a collection of 7,500 posts and 1,260,818 comments with explicit moral judgements. Posters and commenters appear to put a lot of thought and effort into these discussions. Each post contains 381 words on average and each comment contains 57 words on average.



\subsection{Linguistic Analysis of Moral Judgement}
Before we introduce our classifier, we first consider \textbf{RQ1}: what linguistic cues are associated with positive and negative moral judgement?  To answer this question we split comments into two valence classes: positive and negative. The positive class contains comments that are labeled NTA or NAH; the negative class contains comments that are labeled YTA or ESH. Then we use the Allotaxonmeter system, which compares two Zipfian distributions using a scoring function called rank-turbulence divergence, to compare how terms are associated with these valence labels~\cite{dodds2020aAllotaxonometryRankturbulenceDivergence}. In our case, we constructed 1-gram multinomial distributions from each class, which, in the English language, is well known to exhibit a Zipfian distribution~\cite{piantadosi_2014ZipfWordFrequency}.

\begin{table}[t]
\centering
\begin{tabularx}{\linewidth}{@{}XrlX@{}}
    \toprule 
&\textbf{Negative Valence} & \textbf{Positive Valence} &   \\ \midrule
& \tikz[remember picture, overlay]{
    \node[fill=gray!20, anchor=east, rectangle, minimum width=3cm, text width=3cm, inner sep=.5,outer sep=0, align=right]{\phantom{Ty}you};
  }   & 
  \tikz[remember picture, overlay]{
  \node[fill=blue!10, anchor=west, rectangle, minimum width=3cm, text width=3cm, inner sep=.5,outer sep=0, align=left]{to\phantom{Ty}};
  } &
  \\ 
& \tikz[remember picture, overlay]{
    \node[fill=gray!20, anchor=east, rectangle, minimum width=7.5mm, text width=7.5mm, inner sep=.5,outer sep=0, align=right]{quilt};
  }   & 
  \tikz[remember picture, overlay]{
  \node[fill=blue!10, anchor=west, rectangle, minimum width=7.2mm, text width=7.2mm, inner sep=.5,outer sep=0, align=left]{she\phantom{Ty}};
  } &
  \\ 
& \tikz[remember picture, overlay]{
    \node[fill=white, anchor=east, rectangle, minimum width=7.4mm, text width=18.3mm, inner sep=.5,outer sep=0, align=right]{be\phantom{iause}};
    \node[fill=gray!20, anchor=east, rectangle, minimum width=7.3mm, text width=7.3mm, inner sep=.5,outer sep=0, align=right]{cause\phantom{T}};
  }  & 
  \tikz[remember picture, overlay]{
  \node[fill=blue!10, anchor=west, rectangle, minimum width=7.1mm, text width=7.1mm, inner sep=.5,outer sep=0, align=left]{my\phantom{Ty}};
  } &
  \\ 
& \tikz[remember picture, overlay]{
    \node[fill=white, anchor=east, rectangle, minimum width=7.1mm, text width=18.3mm, inner sep=.5,outer sep=0, align=right]{i\phantom{Tynn}};
    \node[fill=gray!20, anchor=east, rectangle, minimum width=7.1mm, text width=7.1mm, inner sep=.5,outer sep=0, align=right]{ntern\phantom{Ty}};
  }  & 
  \tikz[remember picture, overlay]{
  \node[fill=blue!10, anchor=west, rectangle, minimum width=7.0mm, text width=7.0mm, inner sep=.5,outer sep=0, align=left]{cornell\phantom{Ty}};
  } &
  \\ 
& \tikz[remember picture, overlay]{
    \node[fill=gray!20, anchor=east, rectangle, minimum width=6.8mm, text width=6.8mm, inner sep=.5,outer sep=0, align=right]{suck\phantom{T}};
  }  & 
  \tikz[remember picture, overlay]{
  \node[fill=blue!10, anchor=west, rectangle, minimum width=6.6mm, text width=6.6mm, inner sep=.5,outer sep=0, align=left]{they\phantom{Ty}};
  } &
  \\ 
\end{tabularx}
\caption{Rank divergence scores of terms associated with positive and negative moral valences. Color bars indicate the relative contribution of terms, \ie, \textsf{you} contributes about 4 times as much to negative valence as \textsf{quilt}.}
\label{tab:allotax}
\end{table}

\begin{table*}[t]
    \centering
    \begin{tabular}{r| llll}
    \toprule
    \textbf{Method} & \textbf{Accuracy} & \textbf{Precision} & \textbf{Recall} & \textbf{F1} \\
    \midrule
     Doc2Vec Embeddings   & 65.92 $\pm$ 0.04 & 61.22 $\pm$ 0.15 & 13.5 $\pm$ 0.09 & 22.1 $\pm$ 0.09  \\
     BERT Embeddings     & 70.10 $\pm$ 0.07 & 64.28 $\pm$ 0.2 & 36.96 $\pm$ 0.14 & 46.96 $\pm$ 0.08 \\
     Multinomial Na\"ive Bayes & 72.12 $\pm$ 0.17 & 62.58 $\pm$ 0.17 & 55.22 $\pm$ 0.07 & 58.66 $\pm$ 0.05 \\
     Judge-BERT & 89.03 $\pm$ 0.13 & 85.57 $\pm$ 0.18 & 83.48 $\pm$ 0.27 & 84.51 $\pm$ 0.17 \\
     \bottomrule
    \end{tabular}
    \caption{Classification results on the AITA dataset. Results are mean-averages and standard deviations over five-fold cross-validation.}
    \label{tab:binary_classification_results}
\end{table*}

Table~\ref{tab:allotax} shows the terms that contain the largest \textit{divergence contribution} for each valence class. Words with the highest negative valence include \textsf{you}, \textsf{because}, and \textsf{suck} in the top 5, but also \textsf{petty}, \textsf{daughter} and \textsf{jesus} within the top 10 (not shown). Simply put, these are the top words that are used when assigning negative moral judgement. Words associated with positive moral valence consist mainly of functional terms, but also include the names several ivy league schools in the top 50\footnote{A brief survey of these posts reveals that many posters ask if they are wrong to attend one of these schools even though their family member or partner was not admitted.}. 




\subsection{Prediction Model}
Given the dataset, a dataset with textual posts and textual comments labeled with positive or negative moral judgements, our goal is to predict whether an unlabeled comment assigns a positive (NTA or NAH) or negative (YTA or ESH) moral judgement to the user of the post.
It is important to note that this classifier is classifying the judgement of the commenter, not the morality of the poster. 

We define our problem formally as follows.
\paragraph{Problem Definition}
Given a top level comment $C$ with moral judgement $A \in \{+,-\}$ that responded to post $P$ we aim to find a predictive function $f$ such that
\begin{equation}
    f : (C) \rightarrow A
\end{equation}
Formally, this takes the form of a text classification task where class inference denotes the valence of a moral judgement.
The choice of classification model $f$ is not particularly important, but we aim to train a model that performs well and generalizes to other datasets. 
We selected four text classification models for use in the current work:
\begin{itemize}
\item Multinomial Na\"ive Bayes~\cite{kibriya2004multinomial}: Uses word counts to learn a text classification model and has shown success in a wide variety of text classification problems. 
\item Doc2Vec~\cite{le2014distributed}: Create comment-embeddings, which are input into a logistic regression classifier that calculates the class margin.
\item BERT Embeddings~\cite{devlin2018bert}: Uses word embeddings from BERT, which are averaged together and input into a logistic regression classifier that calculates the class margin.
\item Judge-BERT: We fine-tune the BERT-base model using the class labels. Specifically, we added a single dropout layer after BERT's final layer, followed by a final output layer that consists of our two classes. The model is trained using the Adam optimizer and uses the cross entropy loss function. We then trained for 3 epochs as recommended by Devlin et al~\cite{devlin2018bert}.
\end{itemize}

\subsection{Judgement Classification Results}

We evaluate our four classifiers using accuracy, precision, recall, and F1 metrics. In this context a false positive is the instance when the classifier improperly assigns a negative (\ie, asshole) label to a positive judgement. A false negative is the instance when the classifier improperly assigns a positive (\ie, non-asshole) label to a negative judgement. We perform 5-fold cross-validation and, for each metric, report the mean-average and standard deviation over the 5 folds. 


The results in Table~\ref{tab:binary_classification_results} indicate that the Doc2Vec, BERT, and Multinomial Na\"ive Bayes classifiers do not perform particularly well at this task. Fortunately, the fine-tuned Judge-BERT classifier performs relatively well, with an accuracy near 90\% and where type 1 and type 2 errors are relatively similar. Overall, these results indicate that the Judge-BERT classifier is able to accurately classify moral judgements.

\section{Analysis of Moral Judgement}

Using the Judge-BERT classifier, our next tasks are to better understand moral judgement across a variety of online social contexts and to analyze various trends in moral judgement. In order to minimize the transfer-error rate it is important to select subreddit-communities that are similar to the training dataset. In total we chose ten subreddits to explore in our initial analysis. The subreddits we chose can be broken into three main stylistic groups and are briefly described in Table~\ref{tab:subreddits}.

\begin{table}[t]
    \centering
    \begin{tabularx}{\linewidth}{@{}l@{\hspace{1em}}r|X} \toprule
        & \normalsize\textbf{Subreddit} & \normalsize\textbf{Description} \\ \midrule
       \parbox[t]{0mm}{\multirow{1}{*}{\textbf{Advice}}} &  & 
       \multirow{6}{4.3cm}{Users pose questions in a scenario like the AITA dataset and receive advice or feedback on their situation.}\\
        & {/r/relationship\_advice}  & \\
        & {/r/relationships}  & \\
        & {/r/dating\_advice}  & \\
        & {/r/legaladvice}  & \\
        & {/r/dating}  & \\\midrule 
        \parbox[t]{0mm}{\multirow{1}{*}{\textbf{Confessionals}}} &  & 
       \multirow{5}{4.3cm}{Users confess to something that they have been keeping to themselves. Typically, confessions are about something immoral the poster has done.}\\
        & {/r/offmychest} & \\
        & {/r/TrueOffMyChest}  & \\
        & {/r/confessions}  & \\ 
               & & \\
        & & \\ \midrule 
        \parbox[t]{0mm}{\multirow{1}{*}{\textbf{Conversational}}} &  & 
       \multirow{6}{4.8cm}{Users engage in conversations with others to have a simple conversation or to here other opinions in order to change their worldview.}\\

        & {/r/CasualConversation} & \\
        & {/r/changemyview}  & \\
        & & \\
        & & \\
                & & \\
        \bottomrule
        
    \end{tabularx}
    
    \caption{Subreddits used for analysis of moral judgement.}
    \label{tab:subreddits}
\end{table}

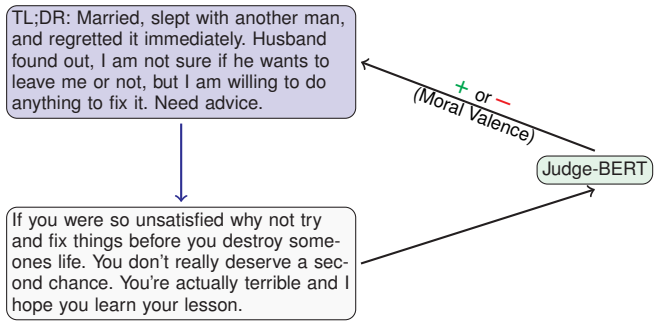
\begin{figure}
    \centering
    \begin{tikzpicture}[font=\sffamily\scriptsize]

\node [examplepostnode, anchor=north] (v1) at (0, 0) {TL;DR: Married, slept with another man, and regretted it immediately. Husband found out, I am not sure if he wants to leave me or not, but I am willing to do anything to fix it. Need advice.};

\node [exampleresponsenode] (v2) at (0,-3.5) {If you were so unsatisfied why not try and fix things before you destroy someones life. You don't really deserve a second chance. You're actually terrible and I hope you learn your lesson.};

\node [genericnode, anchor=north] (v3) at (5.5,-2.0) {Judge-BERT};

\draw [postedge]  (v1) edge  (v2);
\draw[genericedge] (v2.east)  to (v3.south);
\draw[genericedge] (v3.north)  to node[sloped, align=center]{\normalsize{\textcolor{green}{+}} \scriptsize{or} \normalsize{\textcolor{red}{--}}\\ \scriptsize{(Moral  Valence)}} (v1.east);

\end{tikzpicture}
    \caption{A post (in blue) made by a user along with the top response comment (white). The comment is then fed to our Judge-BERT classifier (green) to determine the moral valence of the post.}
    \label{fig:class_example}
\end{figure}

We applied the Judge-BERT classifier to the comments and posts of these ten subreddits. Specifically, given a post and its comment tree we identified the top-level comment with the highest score. This top-rated comment, which has received the most upvotes from the community, is considered to be the one passing judgement on the original poster. As illustrated in Fig.~\ref{fig:class_example}, this top-rated comment is then fed to our classifier and the resulting prediction is used to label the moral valence of the post and poster. It is important to be clear here: we are \textit{not} predicting the moral valence of the comment itself, but rather the top-rated comment is used to pass judgement on the post. 

\nop{
\begin{table}[t]
    \centering
    \begin{tabularx}{\linewidth}{@{}l@{\hspace{1em}}r| XX@{}}
    \toprule
        & \textbf{Subreddit} & \textbf{$\mu^+$ Score} & \textbf{$\mu^-$ Score}\\
    \midrule
       \parbox[t]{0mm}{\multirow{1}{*}{\textbf{Advice}}} &  & &\\
    &/r/relationship\_advice   & 44.32$\pm$1.77 & 23.90$\pm$1.81 \\
&    /r/relationships & 47.19$\pm$1.42 & 23.36$\pm$1.38 \\
&    /r/dating\_advice   & 43.49$\pm$2.51 & 33.45$\pm$3.03 \\
&    /r/legaladvice  & 49.99$\pm$2.38 & 30.07$\pm$2.58 \\
&    /r/dating  & 55.03$\pm$3.45 & 45.09$\pm$4.43 \\ \midrule
\parbox[t]{0mm}{\multirow{1}{*}{\textbf{Confessionals}}} &  & &\\
&         /r/offmychest  & 50.78$\pm$2.13 & 29.67$\pm$2.38 \\ 
&         /r/TrueOffMyChest  & 47.93$\pm$1.40 & 23.95$\pm$1.38 \\ 
&         /r/confessions  & 62.09$\pm$3.95 & 50.92$\pm$5.08 \\ \midrule
\parbox[t]{0mm}{\multirow{1}{*}{\textbf{Conversational}}} &  & &\\
&         /r/CasualConversation  &  46.38$\pm$4.27 & 42.73$\pm$8.71 \\ 
&          /r/changemyview  & 55.17$\pm$3.98 & 56.45$\pm$6.52 \\ 
    \bottomrule
    \end{tabularx}
    \caption{Mean scores and 95\% confidence intervals of posts that were classified as having positive moral valence and negative moral valence. Positive posts are statistically significantly higher than negative posts ($p<0.001$).}
    \label{tab:popularity_intervals}
\end{table}
}

\subsection{Moral Valence and Popularity}

Here we can begin to answer \textbf{RQ2}: Is moral valence correlated with the score of a post? In other words, do posts with positive moral valence score higher or lower than posts with negative moral valence. To answer this question, we extracted all posts and their highest scoring top-level comment from 2018 from each subreddit in Table~\ref{tab:subreddits}.

\begin{figure}[t]
    \centering
    \input{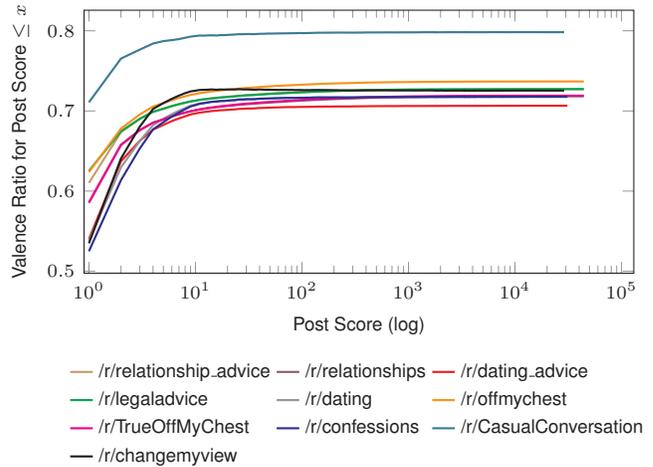}
    \caption{Posts judged to have positive valence as a function of post score. Higher indicates more positive valence. Higher post scores are associated with more positive valence (Mann Whitney $\tau\in[0.395,0.41]$, $p<0.001$ two-tailed, Bonferroni corrected)}.
    \label{fig:popularity_scores}
\end{figure}

Popularity scores on Reddit exhibit a power-law distribution, so the mean-scores and their differences will certainly be misleading. Instead, we plot the ratio of comments judged to be positive against all comments as a function of the post score cumulatively in Fig.~\ref{fig:popularity_scores}. Higher values in the plot indicate more positive valence. The results here are clear: post popularity is associated with positive moral valence. Most of the subreddits appear to have similar characteristics except for /r/CasualConversation, which has a much higher positive valence (on average) than the other subreddits. Mann-Whitney Tests for statistical significance on individual subreddits as well the aggregation of these tests with Bonferroni correction found that posts with positive valence have significantly higher scores than posts with negative valence ($\tau\in[0.395,0.41]$, $p<0.001$ two-tailed). 

We take the additional step to argue that correlation does indeed imply causation in this particular case. Because posts are made before votes are cast, and because the text of a post is (typically) unchanged, and if we assume that scores are causally related to the text of the post, then the causal arrow can only point in one direction, \ie, posts with positive moral valence result in higher scores than posts with negative moral valence. 

These findings appear to conflict with other studies that have shown how negative posts elicit anger and encourage a negative feedback loop on social media~\cite{bebbington2017sky,crockett2017moral}. A further inspection of the posts indicated that posts classified as having positive moral valence often found users expressing that a moral norm had been breached. The difference in our results compared to others may be explained by perceived intent, that is, whether or not the moral violation occurred from an intentional agent towards a vulnerable agent, \cf, dyadic morality~\cite{schein2018theory}. Our inspection of comments expressing negative moral judgement confirms that the perceived intent of the poster is critical to the judgement rendered. These negative judgements typically highlight what the poster did wrong and advise the poster to reflect on their actions (or sometimes simply insult the poster). Conversely, we find that many posts judged to be positive clearly show that the poster is the vulnerable agent in the situation to some other intentional agent. The responses to these posts often display sympathy towards the poster and also outrage towards the other party in the scenario. These instances are perhaps best classified as examples of empathetic anger~\cite{hechler2018difference}, which is anger expressed over the harm that has been done to another. We also note that some of the content labelled to have positive moral valence is simply devoid of a moral scenario. Examples of this can be primarily seen in /r/CasualConversation where the majority of posts are about innocuous topics. 

Another possible explanation for our findings is that users on other online social media sites like Facebook and Twitter are more likely to like and share news headlines that elicit moral outrage; these social signals are then used by the site's algorithms to spread the headline further throughout the site~\cite{brady2020mad,glenski2017rating}. Furthermore, the content of the articles triggering these moral responses often covers current news events throughout the world. Our Reddit dataset, on the other hand, typically deal with personal stories and therefore tend to not have the same in-group/out-group reactions as those found on viral Facebook or Twitter posts. 

\section{Assigning Judgements to Users}

Next we investigate \textbf{RQ3}: Do certain subreddit-communities attract users whose posts are typically classified by more negative or positive moral judgements? To answer this question we need to reconsider our unit of analysis. Rather than assigning moral valence to the individual post, in this analysis we consider the moral valence of the user who committed the post. To do this, we again find all posts and comments of the ten subreddits and find the highest scoring top-level comment; we classify whether that comment is judging the post to have positive or negative moral valence and then tally this for the posting user. 

Of course, users are also able to post comments and sub-comments. So we expand this analysis to include judgements of users from throughout the entire comment tree. Each comment can have zero or more replies each with its own score. So, for each comment we identify the reply with the highest score and classify whether that reply is judging the comment to have positive or negative moral valence, and then tally this for the commenting user. We do this for each comment that has at least one reply at all levels in the comment tree.

By assigning moral valence scores to users we are able to capture all judgements across the ten subreddits and better-understand their behavior. It is important to remember that the classifier classifies the moral valence of text -- with some amount of uncertainty -- not the user specifically. So we emphasize that we do not label users as ``good'' or ``bad'' explicitly; rather, we identify users as having submitted posts and comments that were similar to comments that previously received positive or negative moral judgement. 

\begin{figure}
\centering
    \input{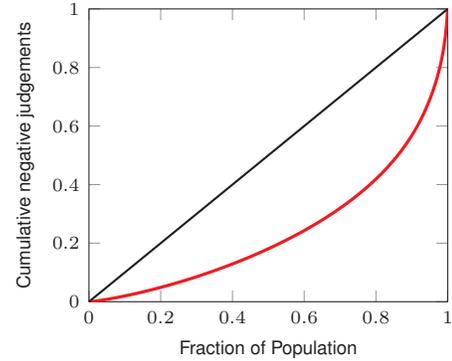}
    \caption{Lorenz curve depicting the judgement inequality among users; Gini coefficient = 0.515}
    \label{fig:lorenz_curve}
\end{figure}

We include only users that were judged at least 50 times. Each user therefore has an associated count of positive and negative judgements. This begs an interesting question: are some users judged more positively or negatively than others? What does that distribution look like? To understand this breakdown we first plot a Lorenz curve in Fig.~\ref{fig:lorenz_curve}. We find that the distribution of moral valence is highly unequal: about 10\% of users receive almost 40\% of the observed negative judgements (Gini coefficient = 0.515). 

This clearly indicates that there are a handful of users that receive the vast majority of negative judgements. To identify those users which receive a statistically significant proportion of negative judgements we perform a one-sided binomial test on each user. Simply put, this test emits a \textit{negativity probability}, \ie, the probability (p-value) that the negativity of a user is not due to chance. 

\begin{figure}
    \centering
    \input{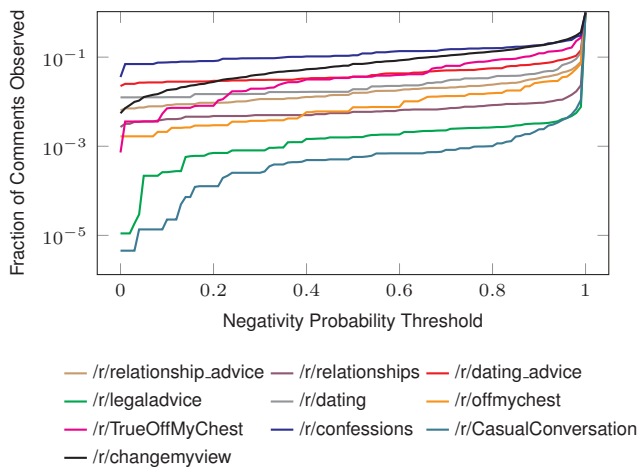}
   \caption{Number of comments (normalized) as a the negativity threshold is raised. As the negativity threshold is raised the fraction of comments revealed tends towards 1. Higher lines indicate a higher concentration of negative users and vice versa.}
    \label{fig:asshole_subreddits_at_50}
\end{figure}

Finally, we can illustrate the membership of each subreddit as a function of users' negativity probability. As expected, Fig.~\ref{fig:asshole_subreddits_at_50} shows that as we increase the negativity threshold from almost certainly negative to uncertainty (from left to right) we begin to increase the fraction of comments observed. These curves therefore indicate the density of comments that are made from negative users (for varying levels of negativity); higher lines (especially on the left) indicate higher concentration of negativity. We find that /r/confessions, /r/changemyview, and /r/TrueOffMyChest contain a higher concentration of comments from more-negative users. On the opposite side of the spectrum, we find that /r/CasualConversation and /r/legaladvice have deep curves, which implies that these communities have fewer negative users than others. 

\subsection{Three Types of Negative Users}

We select our group of users that have a statistically significant negative moral valence as those that were found to have a $p$-value less than 0.05 from our one-tailed binomial test.
Within this group we investigated into their posting habits to determine what types of posts they make to garner such a large number of negative judgements. From our analysis of these users we determined that they fall into three different stylistic groups.

\begin{enumerate}
    \item Explainer: These users will argue that what they did isn't that wrong.
    \item Stubborn Opinion: Users that do not acquiesce to the prevailing opinion of the responders.
    \item Returner: Users that repeatedly post the same situation hoping to elicit more-favorable responses.
\end{enumerate}

The first type of user that we observe is the Explainer. The explainer typically makes a post and receives comments that condemn their immoral actions. In response to this judgement, the explainer will reply to many of the comments in an attempt to convince others that what they did was in fact moral. Often, this only serves to exacerbate the judgements made against them. This then leads to further negative judgements. In fact, we found that many of these users have only made a handful of posts that each have a large number of comments in self-defense. The large number of users that respond to these comments and with negative judgements is similar to the effect of online firestorms~\cite{rost2016digital} but at a scale contained to only an individual post. For these types of posts we also note that some people do come to the defense of the poster, which follows similar findings that people show sympathy after a person has experienced a large amount of outrage \cite{sawaoka2018paradox}.

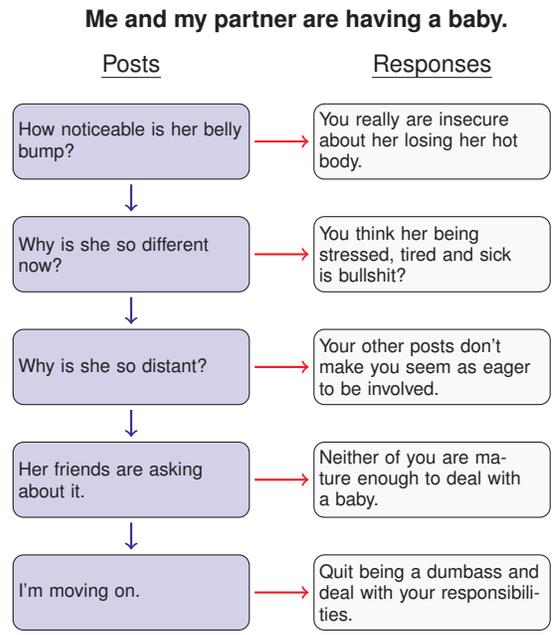
\begin{figure}[t!]
    \centering
    \begin{tikzpicture}[font=\sffamily\scriptsize]

\node [] (v0) at (2.2, 1.6) {\small\textbf{Me and my partner are having a baby.}};

\node[] at (0, 1) {\small\uline{Posts}};
\node[] at (4.0,1) {\small\uline{Responses}};

\node [textnode] (v1) at (0, 0) {How noticeable is her belly bump?};
\node  [textnode] (v2) at (0,-1.5) {Why is she so different now?}; 
\node [textnode] (v3) at (0, -3.0) {Why is she so distant?};

\node [textnode] (v5) at (0, -4.5) {Her friends are asking about it.};
\node [textnode] (v6) at (0, -6.0) {I'm moving on.};

\node[responsenode] (v7) at (4.0,0) {You really are insecure about her losing her hot body.};
\node[responsenode] (v8) at (4.0, -1.5) {You think her being stressed, tired and sick is bullshit?};
\node[responsenode] (v9) at (4.0, -3.0) {Your other posts don't make you seem as eager to be involved.};
\node[responsenode] (v11) at (4.0, -4.5) {Neither of you are mature enough to deal with a baby.};
\node[responsenode] (v12) at (4.0, -6.0) {Quit being a dumbass and deal with your responsibilities.};

\draw [postedge]  (v1) edge  (v2);
\draw [postedge]  (v2) edge (v3);
\draw [postedge]  (v3) edge (v5);

\draw [postedge]  (v5) edge (v6);

\draw [commentedge] (v1) edge (v7);
\draw [commentedge] (v2) edge (v8);
\draw [commentedge] (v3) edge (v9);
\draw [commentedge] (v5) edge (v11);
\draw [commentedge] (v6) edge (v12);

\end{tikzpicture}
    \caption{A diagram showing the posting habits of a Returner. Posts are in the light blue boxes with blue arrows represent the order of posts. An example of a post response is shown in the white box with the red arrow representing the post it came from. Each post is prefaced with the overarching title, "Me and my partner are having a baby." followed by the current update on the situation. The response comments have also been condensed from their full length.}
    \label{fig:returning_asshole}
\end{figure}

The second type of user we observe is the Stubborn Opinion user. These users are similar to but opposite from the Explainers. Rather than trying to change their perspective, the Stubborn Opinion user refuses to acquiesce to the prevailing opinion of the comment thread. For example, users posting to /r/changemyview that do not express a change of opinion despite the efforts and agreement of the commenting users often incur comments casting negative judgement. This back-and-forth sometimes becomes hostile. Many of these conversations end in personal attacks from one of the participants, which has also been shown in previous work on conversations derailing in /r/changemyview~\cite{chang2019trouble}.


The third type of user is the Returner. The returner seeks repeated feedback from Reddit on the same subject. For example, when returners make posts seeking moral judgement, they will often engage in some of the discussion and may even agree with some of the critical responses. Some time later, the user will return and edit their original post or make another post providing an update about their situation. An example of a Returner is illustrated in Figure~\ref{fig:returning_asshole}. In this case, a user continues to request advice after recently impregnating their partner. In these situations responding users often find previous posts on the same topic made by the same user and then use this information and highlight commentary from the previous post to build a stronger case against the user or highlight how the new post is nothing but a thinly-veiled attempt to shine a more-favorable light on their original situation. These attempts usually backfire and result in more negative judgments being cast against the user.

\begin{table}[t]

    \centering
    \begin{tabular}{@{}lll@{}} 
     \multicolumn{3}{c}{/r/relationship\_advice}  \\
    \toprule
    &\textbf{Positive}&\textbf{Negative}\\ \midrule
    \textbf{Male}&53,416&26,281\\
    \textbf{Female}&57,126&20,714\\
    \bottomrule
    \end{tabular}
    
    \vspace{0.5cm}
    
    \begin{tabular}{@{}lll@{}} 
     \multicolumn{3}{c}{/r/relationships}  \\
    \toprule
    &\textbf{Positive}&\textbf{Negative}\\ \midrule
    \textbf{Male}&139,163&74,384\\
    \textbf{Female}&216,190&78,823\\
    \bottomrule
    \end{tabular}
    \caption{Contingency tables showing the number of positive and negative judgements for each self-reported gender. Moral valence has a small association with gender on {/r/relationship\_advice} $\phi=0.07, p<0.01$ and /r/relationships $\phi=0.09, p<0.01$}

\label{tab:relationship-contingency}
\end{table}

\subsection{Gender and Age Analysis}
Our final task investigates \textbf{RQ4}: Are self-reported gender and age descriptions associated with positive or negative moral judgements? Recent studies on this topic have found that gender and moral judgements have a strong association~\cite{reynolds2020man}. Specifically, women are perceived to be victims more often than men and harsher punishments are sought for men. The rates at which men commit crimes tends to be higher than the rates of female crime and society generally views crimes as a moral violation~\cite{choy2017explaining}. If we apply these recent findings to our current research question we expect to find that male users will be judged negatively more often than females. 

This task is not usually available on public social media services because gender and age are not typically revealed, while also allowing for anonymous posting. Fortunately, the posting guidelines of /r/relationships and /r/relationship\_advice requires posters to indicate their age and gender in a structured manner directly in the post title. An example of this can be seen here:
\begin{figure}[h!]
 \centering    
    \includegraphics[width=0.47\textwidth]{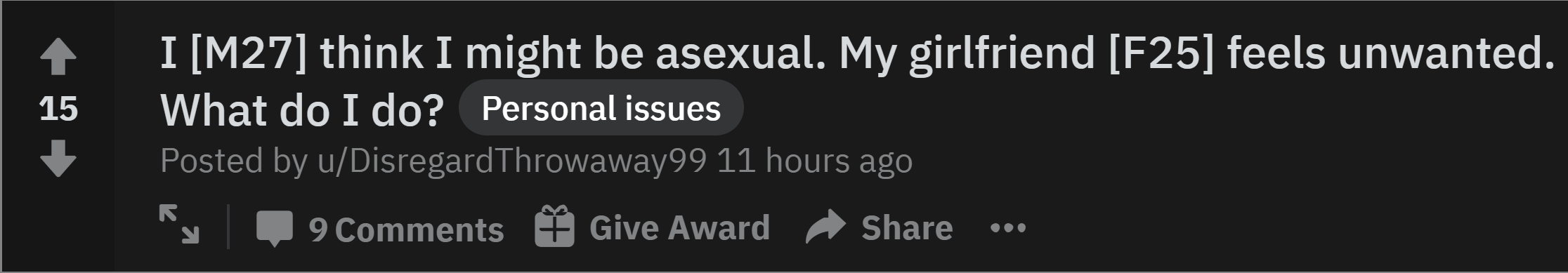} 
\end{figure}

\noindent where the poster uses [M27] to indicate that they identify as male aged 27 years and that their partner [F25] identifies as female aged 25 years. Using these conventions we are able to reliably extract users' self-reported age and gender.

We again apply our Judge-BERT model to assign a moral judgement to the post based on the top-scoring comment. In total we extracted judgements from 508,560 posts on /r/relationships and 157,537 posts on /r/relationship\_advice. Because the posting age of a user may not be above the age of majority, we are careful to only collect data from users that are aged 18 and older via our research ethics guidelines. In general, the age breakdown appears to closely follow Reddit's age demographic. 90\% of posters were between 18-30 years old.

Our first task is to determine if any association exists between moral judgement and gender. To answer this question we performed a $\chi^2$ test of independence. Contingency tables for this test are reported in Table~\ref{tab:relationship-contingency}. The $\chi^2$ test reports a significant association between gender and moral judgement in /r/relationships ($\chi^2 (1, 508,560) = 3874.6, p < .0001$) and in /r/relationship\_advice ($\chi^2 (1, 157,537) = 762.2, p < .0001$). However, the $\chi^2$ test on such large sample sizes usually results in statistical significance; in fact, the $\chi^2$ test tends to find statistical significance for populations greater than 200~\cite{siddiqui2013heuristics}. So we verify this association using $\phi$, which measures the strength of association controlled for population size. In this case, $\phi$ = 0.09 for /r/relationships and $\phi$ = 0.07 for /r/relationship\_advice. These low values indicate that there is only a small association between gender and moral judgement.

\begin{table}[t]
    
     \centering
    \begin{tabular}{l lll }
     \multicolumn{4}{c}{{/r/relationship\_advice}}  \\
    \toprule
        \textbf{Variable} & \textbf{Coefficient} & \textbf{p-value} & \textbf{95\% CI} \\
        \midrule
        \textbf{(Constant)} & -1.1575  & $<$0.001 & (-1.2093, -1.1058) \\ 
        \textbf{Gender} & 0.3076 & $<$0.001 & (0.2806, 0.3241)\\
        \textbf{Age} & 0.0059 & $<$0.001 & (0.0039, 0.0080) \\ 
         \bottomrule
    \end{tabular}
    
    \vspace{0.5cm}
   
    \begin{tabular}{l lll }
     \multicolumn{4}{c}{{/r/relationships}}  \\
    \toprule
        \textbf{Variable} & \textbf{Coefficient} & \textbf{p-value} & \textbf{95\% CI} \\
        \midrule
        \textbf{(Constant)} & -1.0923 & $<$0.001 & (-1.1214, -1.0631) \\
        \textbf{Gender} & 0.3814  & $<$0.001 & (0.3693, 0.3935) \\
        \textbf{Age} & 0.0034  & $<$0.001 & (0.0023, 0.0046) \\

         \bottomrule
    \end{tabular}
    \caption{Results for the logistic regression analysis for both subreddits.}
    \label{tab:relationship_logistic_regression}
\end{table}

\subsection{Logistic Regression Analysis}
Our second task is to determine if gender and age are associated with moral judgement. In other words, are young females, for instance, judged more positively than, say, old males?  
To answer this question, we fit a two-variable logistic regression model where the binary-variable gender is encoded as 0 for female and 1 for male. 

We report the findings from the logistic regressor for each subreddit in Table~\ref{tab:relationship_logistic_regression}. These results indicate that males are judged more negatively than females. Specifically, in /r/relationship\_advice being male is associated with a 35\% increase in receiving a negative judgement. Similarly, in /r/relationships being male is associated with a 46\% increase in receiving a negative judgement.

We also find that age has a relatively small effect on moral judgement; increased age is slightly  correlated with negative judgement. Specifically, in /r/relationship\_advice an increase in age by one year is associated with a 0.59\% increase in receiving a negative judgement. In /r/relationships an increase in age by one year is associated with a 0.34\% increase in receiving a negative judgement.

Simply put, those who are older and those who are male (independently) are statistically more likely to receive negative judgements from Reddit than those who are younger and female. Although gender is much more of a contributing factor than age and neither association is particularly strong.

\section{Conclusions}
In this study, we show that it is possible to learn the language of moral judgements from text taken from /r/AmITheAsshole. We demonstrate that by extracting the labels and fine-tuning a BERT language model we can achieve good performance at predicting whether a user is rendering a positive or negative moral judgement. Using our trained classifier we then analyze a group of subreddits that are thematically similar to /r/AmITheAsshole for underlying trends. Our results showed that users prefer posts that have a positive moral valence rather than a negative moral valence. Another analysis revealed that a small portion of users are judged to have substantially more negative moral valence than others and they tend towards subreddits such as /r/confessions. We also show that these highly negative moral valence users fall into three different types based on their posting habits. Lastly, we demonstrate that age and gender have a minimal effect on whether a user is judged to be have positive or negative moral valence. 

Although the Judge-BERT classifier enabled us to perform a variety of analysis it does pose some limitations. We are unable to verify if the classifier generalizes well to the other subreddits in our study. The test-subreddits do deviate from the types of moral analysis observed in the training data. Moral judgement is not the focus of /r/CasualConversation, for example.  




In the future we hope to implement argument mining in order to gain a better understanding of the reasons for these judgements by extracting the underlying arguments given by users. Other works have done this by extracting rules of thumb through human annotation \cite{forbes2020social} but this limits the ability to perform a large scale analysis. Argument mining has shown success with extracting the persuasive arguments from subreddits like /r/changemyview \cite{dutta2020changing} and would enable us to get a better understanding of moral judgements on social media. This would also allow us to aggregate the underlying themes from these judgements for further analysis.

\section*{Acknowledgements}
We would like to thank Michael Yankoski and Meng Jiang for their help preparing this manuscript. This work is funded by the US Army Research Office (W911NF-17-1-0448) and the US Defense Advanced Research Projects Agency (DARPA W911NF-17-C-0094).

\bibliographystyle{abbrv}
\bibliography{refs}

\end{document}